\documentclass[aps,prx,twocolumn,superscriptaddress]{revtex4-1}

\usepackage{verbatim}
\usepackage{amsmath}
\usepackage[pdftex]{graphicx}
\usepackage[thinspace]{SIunits}
\usepackage[colorlinks=true,urlcolor=blue,linkcolor=blue,citecolor=blue]{hyperref}

\begin{document}

\title{Highly directed emission from self-assembled quantum dots into guided modes in disordered photonic crystal waveguides}

\author{T.~Reichert}\affiliation{Walter Schottky Institut and Physik Department, Technische Universit\"at M\"unchen, Am Coulombwall 4, 85748 Garching, Germany}\affiliation{Nanosystems Initiative Munich, Schellingstra{\ss}e 4, 80799 Mu\"nchen, Germany}
\author{S.~Lichtmannecker}\affiliation{Walter Schottky Institut and Physik Department, Technische Universit\"at M\"unchen, Am Coulombwall 4, 85748 Garching, Germany}\affiliation{Nanosystems Initiative Munich, Schellingstra{\ss}e 4, 80799 Mu\"nchen, Germany}
\author{G.~Reithmaier}\affiliation{Walter Schottky Institut and Physik Department, Technische Universit\"at M\"unchen, Am Coulombwall 4, 85748 Garching, Germany}\affiliation{Nanosystems Initiative Munich, Schellingstra{\ss}e 4, 80799 Mu\"nchen, Germany}
\author{M.~Zeitlmair}\affiliation{Walter Schottky Institut and Physik Department, Technische Universit\"at M\"unchen, Am Coulombwall 4, 85748 Garching, Germany}\affiliation{Nanosystems Initiative Munich, Schellingstra{\ss}e 4, 80799 Mu\"nchen, Germany}
\author{J.~Wembacher}\affiliation{Walter Schottky Institut and Physik Department, Technische Universit\"at M\"unchen, Am Coulombwall 4, 85748 Garching, Germany}\affiliation{Nanosystems Initiative Munich, Schellingstra{\ss}e 4, 80799 Mu\"nchen, Germany}
\author{A.~Rauscher}\affiliation{Walter Schottky Institut and Physik Department, Technische Universit\"at M\"unchen, Am Coulombwall 4, 85748 Garching, Germany}\affiliation{Nanosystems Initiative Munich, Schellingstra{\ss}e 4, 80799 Mu\"nchen, Germany}
\author{M.~Bichler}\affiliation{Walter Schottky Institut and Physik Department, Technische Universit\"at M\"unchen, Am Coulombwall 4, 85748 Garching, Germany}\affiliation{Nanosystems Initiative Munich, Schellingstra{\ss}e 4, 80799 Mu\"nchen, Germany}
\author{K.~M\"uller}\affiliation{Walter Schottky Institut and Physik Department, Technische Universit\"at M\"unchen, Am Coulombwall 4, 85748 Garching, Germany}\affiliation{Nanosystems Initiative Munich, Schellingstra{\ss}e 4, 80799 Mu\"nchen, Germany}\affiliation{E L Ginzton Laboratory, Stanford University, Stanford, CA 94305, USA}
\author{M.~Kaniber}\affiliation{Walter Schottky Institut and Physik Department, Technische Universit\"at M\"unchen, Am Coulombwall 4, 85748 Garching, Germany}\affiliation{Nanosystems Initiative Munich, Schellingstra{\ss}e 4, 80799 Mu\"nchen, Germany}
\author{J.~J.~Finley}\email{finley@wsi.tum.de}\affiliation{Walter Schottky Institut and Physik Department, Technische Universit\"at M\"unchen, Am Coulombwall 4, 85748 Garching, Germany}\affiliation{Nanosystems Initiative Munich, Schellingstra{\ss}e 4, 80799 Mu\"nchen, Germany}

\date{\today}

\begin{abstract}
We explore the dynamics and directionality of spontaneous emission from self-assembled In(Ga)As quantum dots into TE-polarised guided modes in GaAs two-dimensional photonic crystal waveguides. The local group velocity of the guided waveguide mode is probed, with values as low as $\sim 1.5\%\times c$ measured close to the slow-light band edge. By performing complementary continuous wave and time-resolved measurements with detection along, and perpendicular to the waveguide axis we probe the fraction of emission into the waveguide mode ($\beta$-factor). For dots randomly positioned within the unit cell of the photonic crystal waveguide our results show that the emission rate varies from $\geq \unit{1.55}{\nano\second^{-1}}$ close to the slow-light band edge to $\leq \unit{0.25}{\nano\second^{-1}}$ within the two-dimensional photonic bandgap. We measure an average Purcell-factor of $\sim 2\times$ for dots randomly distributed within the waveguide and maximum values of $\beta\sim 90 \%$ close to the slow light band edge. Spatially resolved measurements performed by exciting dots at a well controlled distance $\unit{0-45}{\micro\metre}$ from the waveguide facet highlight the impact of disorder on the slow-light dispersion. Although disorder broadens the spectral width of the slow light region of the waveguide dispersion from $\delta E_{d}\leq\unit{0.5}{\milli\electronvolt}$ to $>\unit{6}{\milli\electronvolt}$, we find that emission is nevertheless primarily directed into propagating waveguide modes. The ability to control the rate and directionality of emission from isolated quantum emitters by placing them in a tailored photonic environment provides much promise for the use of slow-light phenomena to realise efficient single photon sources for quantum optics in a highly integrated setting.
\end{abstract}

\maketitle


\section{\label{sec:section1}Introduction}

Many of the existing proposals for optically based quantum information technologies rely on the availability of efficient sources of single photons \cite{Knill2001,Claudon2010,Reimer2012} and the ability to enhance the strength of light-matter interactions to a level where few photon non-linearities appear in the optical response \cite{O-Brien2009Photonic}.  To date, such quantum optical non-linearities have been demonstrated for several free-space and cavity-QED systems including atoms in high finesse optical resonators \cite{Turchette1995,Nogues1999}, semiconductor quantum dots (QDs) embedded within high-Q (Q = quality factor) solid state nano-cavities \cite{Englund2007Cav_Ref,Fushman2008,Englund2010,Volz2012,Englund2012} and individual dye molecules subject to polychromatic excitation \cite{Hwang2009}. In the light of these impressive demonstrations, several groups have already turned their attention to integrated geometries \cite{Honjo2004,Takesue2005,O-Brien2009Photonic} whereby cavities, waveguides and quantum emitters can be combined on the same chip to realise new types of quantum sources \cite{Munoz2014}. High-$Q$ photonic crystal (PhC) defect cavities can be readily fabricated next to waveguides to effectively direct quantum light into propagating modes on a chip \cite{Yao2009Cav-WG}. However, in-situ frequency control is required to tune the QD-emitter and cavity mode into resonance. In contrast, a broadband spontaneous emission rate enhancement can be achieved using PhC waveguides close to the low group velocity (slow light) regions of the dispersion relation for TE-guided modes \cite{Rao2007Finite,Hughes2004, Viasnoff2005, Lecamp2007,Vlasov2005} with measured coupling efficiencies of the emission to the waveguide mode approaching unity \cite{Lund-Hansen2008,Thyrrestrup2010,Dewhurst2010,Hoang2012}.  Moreover, a recent theoretical proposal \cite{Gao2008} has indicated that the enhanced light-matter interaction close to slow light modes in PhC waveguides may become sufficiently strong such as to result in single photon non-linearities. However, the low group velocity region of the propagating mode is inevitably impacted by disorder effects that can result in Anderson localisation close to bandedges \cite{Gao2013,Savona2011,Huisman2012,Lagendijk2009}, potentially hindering the practical use of slow-light phenomena.

In this paper we combine continuous wave (CW) and time-resolved photoluminescence (PL) spectroscopy to probe the coupling of QDs randomly distributed throughout a PhC W1 waveguide to the TE-polarised guided modes. We measure the local group velocity at specific points within the waveguide dispersion, obtaining values as low as $\sim 1.5\% \times c$ close to the bandedge. This enables us to directly correlate the measured local spontaneous emission rate with the spectrum of the radiation detected along two orthogonal axes; parallel to the waveguide axis and normal to the plane of the two-dimensional (2D) PhC.  Our results show that the average spontaneous emission rate varies from $\geq \unit{1.55}{\nano\second^{-1}}$ for dots emitting close to the slow-light region of the waveguide dispersion to $\leq \unit{0.25}{\nano\second^{-1}}$ within the 2D photonic bandgap. For dots randomly positioned within the unit cell of the PhC waveguide we measure a position averaged Purcell-factor up to $\sim2\times$ and spontaneous emission coupling factors into the guided waveguide mode up to $\beta \sim 90 \%$. Finally, spatially resolved measurements directly elucidate the impact of fabrication disorder on the slow light edge of the dispersion relation. We observe pronounced optical localisation for random positions along the photonic crystal waveguide. The slow light waveguide mode band edge is fluctuating over an energy interval $\delta E_{d} = \unit{6}{\milli\electronvolt}$ due to the presence of disorder.


\section{\label{sec:section2}Fabrication \& Experimental Setup}

The sample investigated was grown using molecular beam epitaxy on a $\unit{350}{\micro\metre}$ thick [100] GaAs wafer. Growth began with a $\unit{800}{\nano\metre}$ thick sacrificial layer of Al$_{0.8}$Ga$_{0.2}$As grown on a $\unit{300}{\nano\metre}$ GaAs buffer, followed by an $\unit{150}{\nano\metre}$ thick nominally undoped GaAs waveguide that contained a single layer of In$_{0.5}$Ga$_{0.5}$As QDs at its midpoint.  The growth conditions used for the QD layer are known to produce dots with an areal density $\rho_D$ \mbox{$\sim 50$ $\mu$m$^{-2}$}, emitting over the spectral range $\unit{1.24-1.40}{\electronvolt}$. After growth, a hexagonal lattice of air holes was defined in a ZEP $520$-A soft mask and deeply etched using a SiCl$_4$ based inductively coupled plasma to form a 2D PhC. By omitting a single row of air holes in the PhC lattice we established a $W1$ waveguide that was subsequently cleaved to gain optical access via the side facet.  As a final step the AlGaAs layer was selectively removed with hydrofluoric acid to establish a free standing membrane.

After fabrication the sample was cooled to $T = \unit{12}{\kelvin}$ in a liquid He flow-cryostat for optical studies using a two-axis confocal microscope that facilitates the study of the optical response both perpendicular and parallel to the sample surface. The QDs were excited with a pulsed laser diode emitting at $\unit{1.9}{\electronvolt}$ ($\unit{80}{\mega\hertz}$ repetition frequency, $\unit{70}{\pico\second}$ pulse duration, Pico Quant, model P-650) along an axis normal to the plane of the waveguide. The signal was detected either via the same $50\times$ objective (NA$ = 0.42$), perpendicular to the plane of the PhC and waveguide axis, or from the side along the waveguide axis using a second $100 \times$ (NA$ = 0.5$) objective. The diameter of the excitation spot was measured to be $\unit{1.3}{\micro\metre}$ such that $\sim50$ QDs are excited directly. We spectrally dispersed the QD emission using a $\unit{0.5}{\metre}$ imaging monochromator and detected with a liquid nitrogen cooled CCD camera. For time-resolved measurements a Si-avalanche photodiode was used, providing a temporal resolution of $\unit{350}{\pico\second}$.

\begin{figure}
\includegraphics{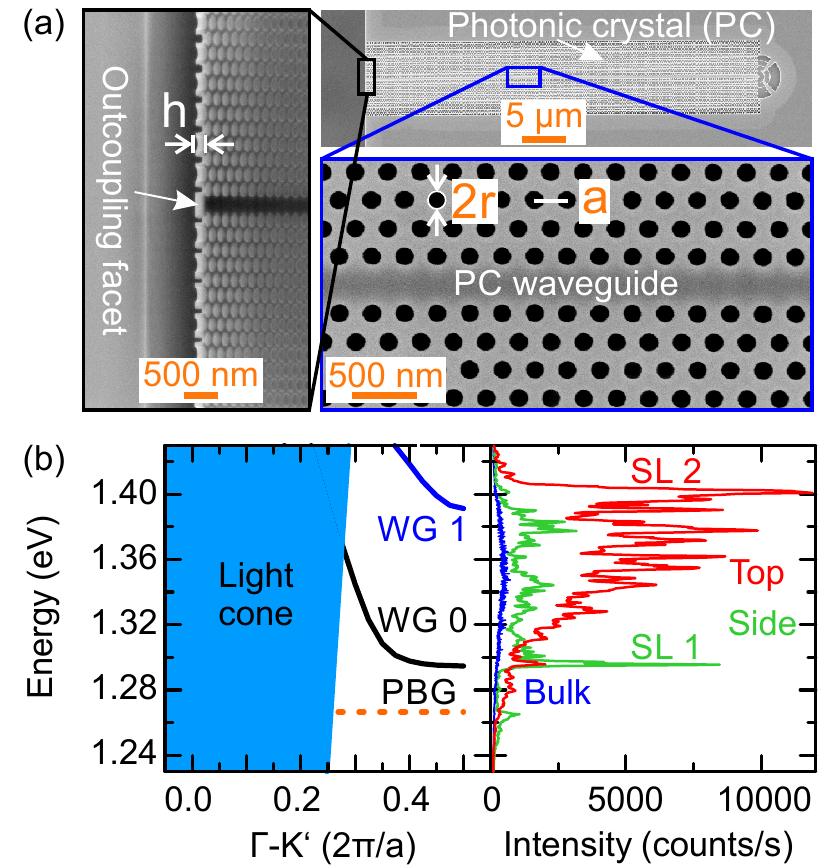}
\caption{ (a) Selected SEM images of one half of the cleaved W1 PhC waveguide, nominally identical to the sample used for the spectroscopic studies. The leftmost image shows the cleaved facet with the under etched membrane, tilted at an angle of $\unit{60}{\degree}$. The rightmost images show planar views. From these images the air hole radius $r = \unit{79 \pm 2}{\nano\metre}$, lattice constant $a = \unit{250 \pm 2}{\nano\metre}$ and the slab thickness $h = \unit{150 \pm 5}{\nano\metre}$ were determined. (b) Left panel: Photonic bandstructure calculation for the structural and geometrical parameters extracted from the SEM images shown in (a). The black and blue solid lines show the $0^{th}$ and $1^{st}$ order waveguide modes, respectively. The orange dashed line marks the upper edge of the 2D photonic band and the light blue shaded region represents the free space light cone. In the rightmost panel we present typical $\micro$-PL measurements of the investigated structure excited at the position of the waveguide from the top (see text) and detected from the side (green) and top (red), respectively. The blue spectrum shows typical QD emission recorded from the top using identical conditions from the unprocessed region of the sample for comparison.
 \label{figure1}}
\end{figure}

\mbox{Figure \ref{figure1} (a)} shows selected scanning electron microscope (SEM) images of a structure that is nominally identical to the one used for optical studies.

From such SEM images we determined the air hole radius to be $r/a = 0.315 \pm 0.005$, where $a=\unit{250 \pm 2}{\nano\metre}$ is the nominal periodicity of the PhC, the slab thickness $h = \unit{150 \pm 5}{\nano\metre}$ and the waveguide length $l = \unit{45 \pm 0.5}{\micro\metre}$. Using the extracted geometrical parameters and the refractive index for GaAs of $n_{GaAs} = 3.5$, we performed three-dimensional calculations of the photonic bandstructure \cite{RSoft2014}.  Selected examples of such calculations are presented in the leftmost panel of \mbox{Fig. \ref{figure1} (b)} that shows the dispersion for TE-like modes along the $\Gamma$-$K'$ direction in the first Brillouin zone \cite{Johnson2000, Dorfner2008}. The fundamental ($0^{th}$) and first ($1^{st}$) order waveguide modes are labelled $WG$ 0 (black line) and $WG$ 1 (blue line), respectively, while the light blue shaded region marks the light cone and the orange dashed line marks the position of the edge of the 2D photonic bandgap. Clearly, the fundamental waveguide mode is expected to be guided in the spectral range $E = \unit{1.296}{\electronvolt} - \unit{1.361}{\electronvolt}$, below the light-line and the $1^{st}$ order waveguide mode from $E = \unit{1.391}{\electronvolt} - \unit{1.439}{\electronvolt}$, within the 2D photonic bandgap.

The rightmost panel of \mbox{Fig. \ref{figure1} (b)} shows typical QD PL spectra recorded by exciting a specific position on the waveguide $\unit{10}{\micro\metre}$ from the cleaved facet and detecting emission either from the facet (green spectrum), or normal to the plane of the waveguide at the excitation position (red spectrum). The blue spectrum shows a typical emission spectrum recorded using nominally identical conditions from the unprocessed region of the sample, with a near featureless form reflecting the comparatively large number of QDs excited. In contrast, the spectra recorded from the waveguide exhibit a narrow peak close to the slow light region of the fundamental waveguide mode at $E^{SL}_1 = \unit{1.296}{\electronvolt}$, labelled $SL_1$ in \mbox{Fig. \ref{figure1} (b)}. The energy of the peak $SL_1$ is in excellent agreement with our photonic bandstructure simulations and the feature is observed for both top- and side- detection geometries.  Similarly, a weaker additional feature, labelled $SL_2$ in \mbox{Fig. \ref{figure1} (b)}, is observed only for the top detection geometry at $E^{SL}_2 = \unit{1.391}{\electronvolt}$.  We identify $SL_2$ as arising from the $1^{st}$ order waveguide mode, its absence in the side detection geometry is most probably a result of the higher propagation losses of the higher energy waveguide mode.  The characteristic form of the emission spectrum clearly indicates that the tailored photonic mode density experienced by QDs strongly modifies the directionality of the spontaneous emission, as expected.


\section{\label{sec:section3}Results and Discussion}

In this section we present a detailed study of the modified quantum dot emission properties, the decreased group velocity and the impact of disorder on the slow light cut-off energy. To gain insight into the quantum dot emission dynamics we performed time-resolved PL measurements to directly probe the local photonic mode density experienced by the dots within the waveguide. The modified photonic mode density is expected to strongly influence the radiative decay rate according to the effective Purcell factor which is given by:
\begin{equation}
F_P = (3\pi c^3)/(A_{eff}\omega_{QD}^2\epsilon^{3/2})\times(1/v_g),
\label{Purcell}
\end{equation}
where $A_{eff}$ is the effective mode area and $v_g = \hbar^{-1}(dE/dk)$ the local group velocity of the waveguide mode \cite{Rao2007Finite}. The enhanced density of propagating modes close to the slow light regions of the waveguide dispersion are expected to influence the directionality of the spontaneous emission.  We measured the frequency dependence of the spontaneous emission rate spanning the energy range between $\unit{1.225}{\electronvolt}$ and $\unit{1.395}{\electronvolt}$, overlapping with the 2D photonic bandgap and the waveguide modes. Hereby, we used the spectrometer as a spectral filter with a bandpass of $\unit{0.5}{\milli\electronvolt}$ and recorded decay transients in steps of $\unit{2}{\milli\electronvolt}$ via the waveguide facet with the excitation laser positioned on the waveguide $\unit{10}{\micro\meter}$ away from the facet. 

\begin{figure}
\includegraphics[scale=1.00]{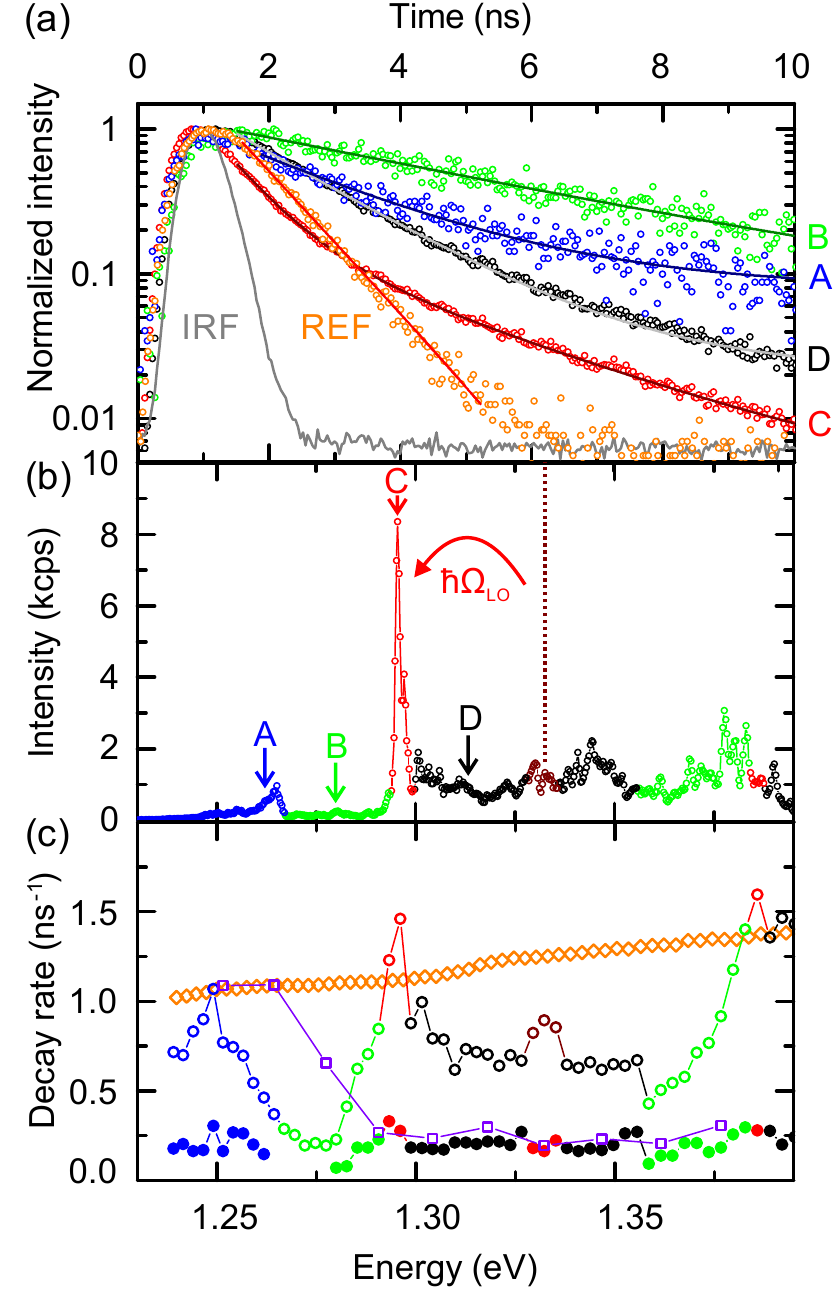}%
\caption{(a) Comparison of a typical decay transient recorded at a detection energy of $\sim \unit{1.3}{\electronvolt}$ from dots in the unpatterned region (REF) of the samples and selected decay transients ($A,B,C,D$) recorded at the spectral positions indicated in panel-b using the measurement geometry described in the text (IRF - Instrument Response Function). (b) Corresponding $\mu$-PL spectrum detected from the side with top excitation $\unit{10}{\micro\metre}$ away from the facet. The slab band region is marked in blue, the photonic bandgaps in green, the slow light regions of the $0^{th}$ and $1^{st}$ order mode in red, the fast guided mode region in black and the phonon coupled slow light region in brown. (c) Spontaneous emission rate of the QDs in the PhC waveguide as a function of the emission energy extracted from the decay transients as measured in panel-a. For comparison, the decay rate from QDs in the unprocessed region of the sample and on the PhC membrane away from the waveguide is shown by the orange and purple squares respectively.
 \label{figure2}}
\end{figure}

Typical intrinsic decay rates for QDs within the unpatterned region of the GaAs membrane lie in the range $\unit{1 - 1.25}{\nano\second^{-1}}$, increasing monotonically toward higher emission energies presumably as a consequence of the large coherence volume in more strongly confined dots.  In strong contrast, the spectral evolution of the measured emission rate from dots within the waveguide exhibits considerably more complex decay dynamics with a much richer spectral dependance. Typical representative data are presented in \mbox{Fig. \ref{figure2} (a)} for the excitation spot positioned $\sim \unit{10}{\micro\metre}$ from the waveguide facet and a range of different detection energies; $A$ - outside the photonic bandgap (blue trace - \mbox{Fig. \ref{figure2} (a)}), $B$ - within the photonic bandgap below the slow light region of the waveguide dispersion (green trace - \mbox{Fig. \ref{figure2} (a)}), $C$ - at the slow light edge of the fundamental waveguide mode dispersion (red trace - \mbox{Fig. \ref{figure2} (a)}) and $D$ - close to the guided, fast light region of the fundamental waveguide dispersion (black trace - \mbox{Fig. \ref{figure2} (a)}).  Careful examination of the data presented in \mbox{Fig. \ref{figure2}} shows that all decay transients can be well described by either mono- or bi-exponential fits of the form  $I(t) = I_1\times exp(-t/\tau_1) + I_2\times exp(-t/\tau_2)$, respectively ($I_2 =0 $ for mono-exponential fits).

We now continue to discuss the form of these decay transients: Resonant with the slab mode continuum ($A$) we observe a bi-exponential decay from which we obtain a fast decay rate of $\Gamma^{Slab}_1 = \unit{0.5 \pm 0.02}{\nano\second}^{-1}$ and a slow component $\Gamma^{Slab}_2 = \unit{0.25 \pm 0.02}{\nano\second}^{-1}$. Inside the photonic bandgap ($B$), however, we measure a mono-exponential decay and observe a significant decrease of the spontaneous emission rate to $\Gamma^{PBG} = \unit{0.2 \pm 0.02}{\nano\second}^{-1}$. In contrast, resonant with the slow light region of the fundamental waveguide mode ($C$) we clearly observe again a bi-exponential decay transient with a fast component $\Gamma^{SL}_1 = \unit{1.55 \pm 0.3}{\nano\second}^{-1}$ and a slow component $\Gamma^{SL}_2 = \unit{0.37 \pm 0.02}{\nano\second}^{-1}$. Finally, in the fast guided mode regime ($D$) we also find that a bi-exponential transient best accounts for the observed dynamics, extracting a fast decay rate of $\Gamma^{FL}_1 = \unit{0.75 \pm 0.05}{\nano\second}^{-1}$ and a slow decay rate $\Gamma^{FL}_2 = \unit{0.17 \pm 0.01}{\nano\second}^{-1}$. The corresponding $\mu$-PL spectrum recorded from the side whilst exciting via the top is plotted in figure \mbox{Fig. \ref{figure2} (b)} for comparison. In \mbox{Fig. \ref{figure2} (c)}, we present the extracted QD spontaneous emission decay rate as a function of energy between $\unit{1.225}{\electronvolt}$ and $\unit{1.395}{\electronvolt}$ in steps of $\unit{2}{\milli\electronvolt}$.  The open orange diamonds represent reference data obtained from dots outside a tailored photonic environment and the color coded circles show the various decay rates measured from the W1 PhC waveguide.  Whenever biexponential decay transients were observed we plot the high and low decay rates in \mbox{Fig. \ref{figure2} (c)} by open and filled circles, respectively. 
By comparing the measured QD decay rate inside the waveguide at the slow light edge at $E^{SL}_{1} = \unit{1.296}{\electronvolt}$ with the reference decay rates in the unpatterned region of the sample at the same energy, we determine average Purcell-factors between $F_{P} = 1 - 2$ for dots randomly positioned within the PhC unit cell, in good accord with expectations in the literature \cite{Laucht2012Ensemble}. For the $1^{st}$ order waveguide mode, we observe a qualitatively similar behaviour, however, far less pronounced, which we attribute to the lower group index as compared to the fundamental mode. Besides the maximum in the decay rate at the slow light edge of the fundamental mode at $E^{SL}_{1} = \unit{1.296}{\electronvolt}$, another weak peak is observed at $\unit{1.332}{\electronvolt}$. The energy separation between these two features is very close to the GaAs longitudinal optical phonon energy ($\hbar\omega_{LO} = \unit{36.6}{\milli\electronvolt}$), indicative of a phonon assisted QD-decay mechanism via the slow light mode \cite{Hohenester2009}. To estimate the fraction of photons coupled to the PhC waveguide mode ($\beta_{\Gamma}$-factor), we also measured the spontaneous emission rate for emission into the photonic bandgap. The result is presented by the purple squares in \mbox{Fig. \ref{figure2} (c)} showing that typical decay rates for dots emitting into the photonic bandgap are $\Gamma^{PBG} = \unit{0.22 \pm 0.02}{\nano\second}^{-1}$. 
From the QD decay rates at the slow light edge and the rates of QDs emitting into the photonic bandgap at the same energy we estimated the single mode spontaneous emission coupling factor $\beta_{\Gamma}$ using
\begin{equation}
\beta_{\Gamma}=\frac{\Gamma_{WG}}{\Gamma_{WG}+\Gamma_{int}}
\label{beta}
\end{equation}
Here, ${\Gamma_{WG}}$ is the QD decay rate into WG modes and ${\Gamma_{int}}$ is the intrinsic emission decay rate in the photonic bandgap \cite{Laucht2012Single}. Using the measured values of $\Gamma^{SL}_1 = \unit{1.55 \pm 0.3}{\nano\second}^{-1}$ and $\Gamma^{PBG} = \unit{0.2 \pm 0.02}{\nano\second}^{-1}$, we estimate $\beta_{\Gamma}= 89 \pm 4 \%$, in good agreement with previously published work \cite{Laucht2012Single,Lecamp2007,Rao2007Beta,Thyrrestrup2010}. 


\begin{figure}
\includegraphics[scale=1.00]{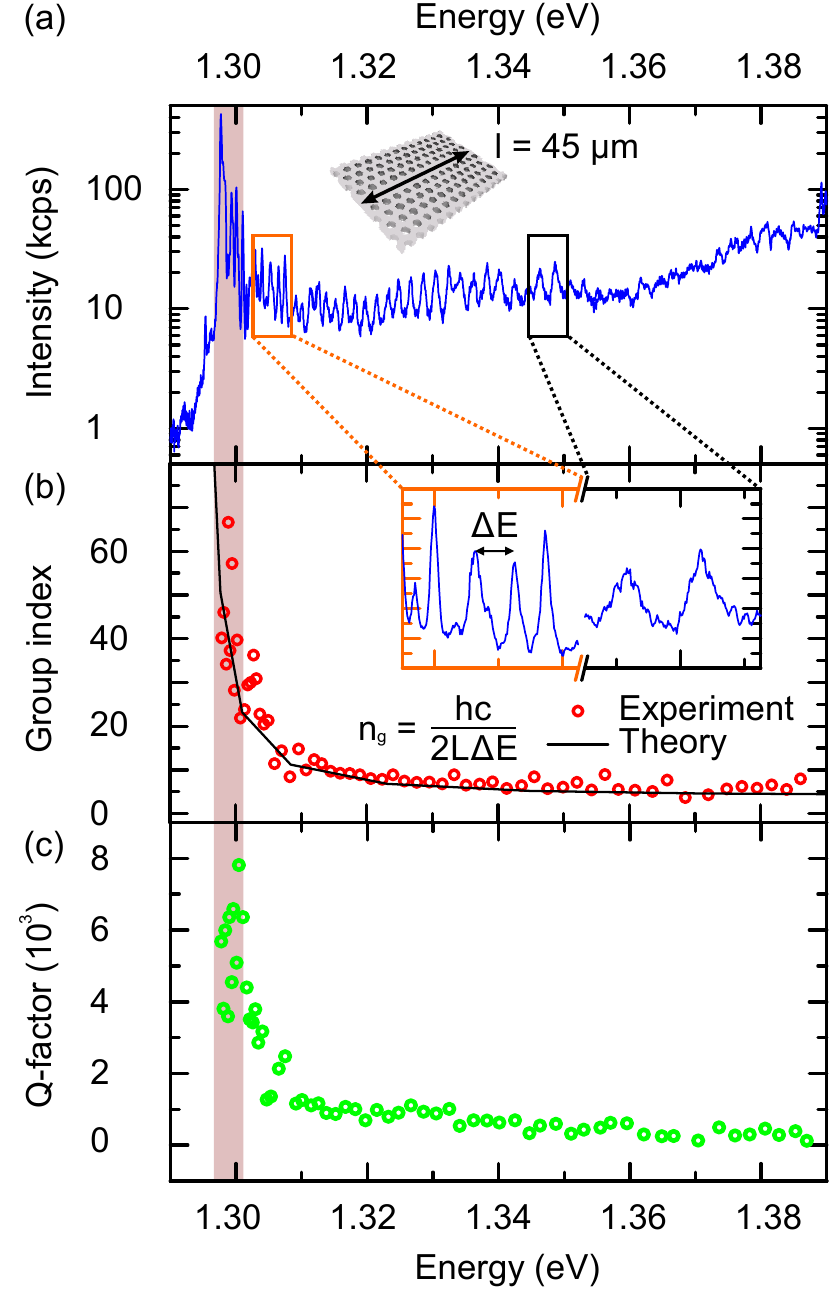}%
\caption{(a) High power ($\unit{30}{\micro\watt}/\mu$m$^{2}$) $\micro$-PL spectrum, detected from the side facet when exciting from the top $\unit{10}{\micro\metre}$ from the facet. (b) Group index as a function of energy; The red circles represent the group index extracted from measured data and the black solid line is the group index derived from the photonic bandstructure simulations presented in \mbox{Fig. \ref{figure1} (b)}; the inset shows a zoom of (a). (c) $Q$-factor extracted from (a) as a function of energy.
 \label{figure3}}
\end{figure}

Purcell factors up to $\sim 30$ have been theoretically predicted to be within reach for a group index as high as $c/v_g \sim 150$, corresponding to wave vectors close to the slow light region of the waveguide mode ($k_{\Gamma-K} \sim 0.47 \pi/a$) \cite{Rao2007Beta}. However, in experiments an ideal spatial location of the emitter within the extended unit cell of the waveguide is crucial to reach these large values of $F_{P}$ \cite{Rao2007Beta}. In order to estimate the expected Purcell factor of an ideally coupled QD emitting at $E^{SL}_{1} = \unit{1.296 \pm 0.001}{\electronvolt}$ we measured the group index of the propagating waveguide mode \cite{Hoang2012} in PL measurements. Therefore, we excited with a high pump fluence ($\unit{30}{\micro\watt}/\mu$m$^{2}$), far above the QD s-shell saturation. Under such excitation conditions, the finite length of the PhC waveguide ($l = \unit{45}{\micro\metre}$) results in the appearance of clear Fabry-Perot oscillations in the waveguide emission.  The local spacing of neighbouring Fabry-Perot maxima are $\Delta E = {hc/2n_{g}L}$, where $n_{g}$ is the mode group index and $L$ is the length of the PhC waveguide. In figure \mbox{Fig. \ref{figure3} (a)} we present a typical high-power $\micro$-PL spectrum, detected from the side facet while exciting from the top $\unit{10}{\micro\metre}$ away from the facet. Fabry-Perot oscillations are clearly observed with a continuous reduction of the mode spacing (from $\unit{2.5}{\milli\electronvolt}$ down to $\unit{0.25}{\milli\electronvolt}$) when approaching the slow light edge (red shaded region) from the high energy side, reflecting the smaller group velocity when approaching the bandedge \cite{Hoang2012}. From the data presented in \mbox{Fig. \ref{figure3}} we calculated the group index $n_{g}$ using
\begin{equation}
n_{g}=\frac{hc}{2L\Delta E}
\label{groupindex}
\end{equation}
In \mbox{Fig. \ref{figure3} (b)} we present the extracted group index $n_g$ as function of energy.  The group index clearly rises from $\sim 5$ to $\geq 70$ when approaching the slow light edge of $WG$ 0. From this we conclude that photons at the slow light edge propagate along the waveguide at only $1.5\%$ of the speed of light in vacuum. The calculated values extracted from the Fabry-Perot resonances are in very good agreement with the theoretical values obtained from the photonic bandstructure simulation (solid black line) confirming the accuracy of these simulations. Simultaneously, the Q-factor of the Fabry-Perot resonances increases from a few hundred in the fast light region of the waveguide dispersion up to $\sim 8000$ close to the slow light edge as shown in \mbox{Fig. \ref{figure3} (c)}, reflecting the enhanced WG losses when approaching the light line.


\begin{figure*}
\includegraphics[scale=1.00]{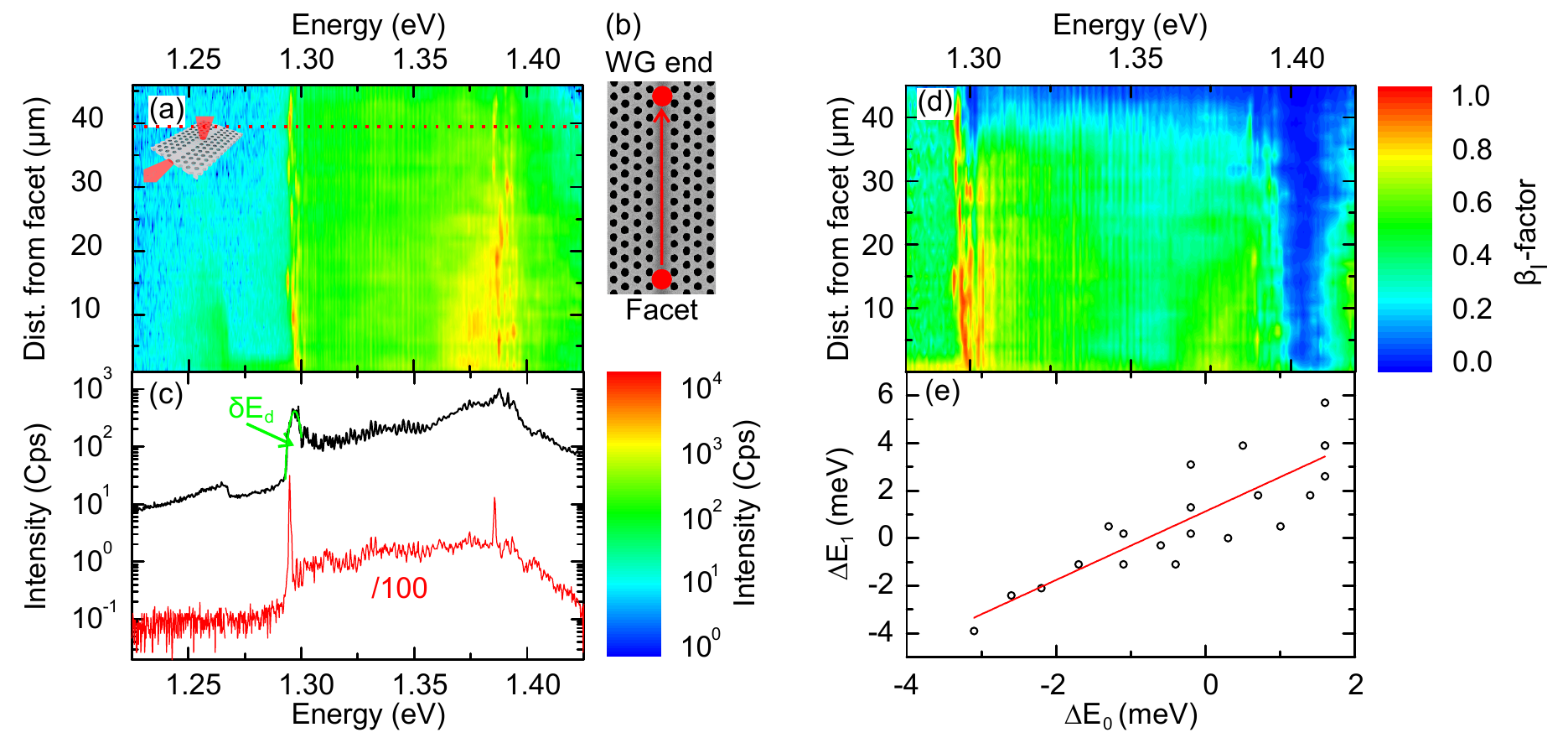}
\caption{(a) Spectrum recorded via the side channel as a function of the excitation position along the waveguide. (b) Top view SEM image from a waveguide identical to the measured one, illustrating the excitation spot scan. (c) Averaged spectrum of (a) (black curve)  and a single spectrum when exciting $\unit{40}{\micro\metre}$ away from the facet (red curve, position is indicated by the red dotted line in (a)). The slow light disorder window width is $\delta E_{d} = \unit{6}{\milli\electronvolt}$. (d) Extracted $\beta_{I}$-factor as a function of position and energy. (e) Disorder induced mode peak deviation of mode $1$ from average as a function of mode peak deviation of mode $0$ from average.
 \label{figure4}}
\end{figure*}

We continue to explore the impact of disorder on the guided modes close to the slow light edge of the waveguide mode. \mbox{Fig. \ref{figure4} (a)} shows a false colour image of PL spectra detected from the \emph{side facet} when moving the excitation spot along the waveguide axis from the facet in steps of $\unit{1}{\micro\metre}$ (see \mbox{Fig. \ref{figure4} (b)}). The spectra obtained clearly reveal a series of localised modes close to the slow light edge $E^{SL}_1 \sim \unit{1.296}{\electronvolt}$, the energy at which fluctuations occur as the excitation spot is shifted along the waveguide.  We identify such features as being due to disorder induced localised modes close to the slow light edge \cite{Gao2013, Savona2011}. We note that the disorder induced cavity mode Q-factors ($\sim6000$ to $\geq8000$) seem mainly to be limited by the in-plane optical confinement since photons still couple \emph{primarily} to propagating waveguide modes, as demonstrated by the fact that we observe them most prominently in the side detection geometry. To quantitatively estimate the impact of disorder on the slow light cut-off energy, we compare in figure \mbox{Fig. \ref{figure4} (c)} the averaged spectrum of all positions shown in \mbox{Fig. \ref{figure4} (a)} (black curve) with a single spectrum detected from the side when exciting $\unit{40}{\micro\metre}$ away from the facet (red curve). For an individual spectrum  we observe a sharp resonance close to the energy of the slow light edge at $E^{SL}_1 = \unit{1.296}{\electronvolt}$, similar in form to many of the spectra at different positions along the waveguide. In strong contrast, in the position averaged spectrum we observe a broadened peak at $\left\langle E^0_d\right\rangle=\unit{1.297}{\electronvolt}$ with a full width at half maximum (FWHM) $\delta E_{d} = \unit{6}{\milli\electronvolt}$. This disorder bandwidth provides a measure of fabrication imperfections along the complete $\unit{45}{\micro\metre}$ length of the PhC waveguide. Furthermore, we investigated the impact of disorder on the $\beta$-factor by recording both side and top detection signals \emph{simultaneously}. We excite at a position distant $x$ from the facet (data not shown) and define a quantity $\beta_I(\omega,x)$:
\begin{equation}
\begin{split}
\beta_{I}(\omega,x)=\frac{\eta_{side} \cdot I_{side}(\omega,x)}{\eta_{side} \cdot I_{side}(\omega,x)+ \eta_{top} \cdot I_{top}(\omega,x)} = \\ {\left(\frac{\eta_{top}}{\eta{side}}\cdot\frac{I_{top}(\omega,x)}{I_{side}(\omega,x)} + 1\right)}^{1/2},
\label{beta_I}
\end{split}
\end{equation}
where $I_{side}(\omega,x)$ and $I_{top}(\omega,x)$ are the PL intensities detected from the side and top, respectively, and $\eta_{side}$ and $\eta_{top}$ are collection efficiencies in these two detection geometries. We assume $\eta_{top}$ and $\eta_{side}$ remain constant during the experiment and obtain $\eta_{top}/\eta_{side}$ $\sim0.22\pm0.07$.  Using this result we obtain $\beta_{I}(\omega,x)$-factor using \mbox{Eqn. \ref{beta_I}} and plot the result in \mbox{Fig. \ref{figure4} (d)}. We observe a region around $\sim\unit{1.3}{\electronvolt}$ with $\beta_I$-factors as high as $90\pm5\%$ which we identify to be the slow light region of $WG$ 0. Along the complete waveguide we identify spatially localised hot spots with remarkably high $\beta_I$-values, demonstrating that those spatial positions can be used to efficiently in-couple light into propagating waveguide modes. In the fast-guided region of the fundamental-mode we observe moderate $\beta_{I}$-factors of $60 \pm 10 \%$, which slightly decrease as the excitation spot moves away from the facet (green region on \mbox{Fig. \ref{figure4} (d)}). In contrast, we identify decreased $\beta_{I}$-factors of $10 \pm 5 \%$ close to $WG$ 1 around $\unit{1.40}{\electronvolt}$ which we attribute to the pronounced losses due to scattering into out-of-plane modes above the light line. Finally, we investigate if the influence of structural disorder impacts simultaneously both waveguide modes. Therefore, we define the energetic separation $\Delta E_0 = \left|E^0_{dis}-\left\langle E^0_d \right\rangle\right|$ ($\Delta E_1 = \left|E^1_{dis}-\left\langle E^1_d\right\rangle\right|$) of a disorder induced localized state $E^{0/1}_{dis}$ with respect to its according average $\left\langle E^0_d \right\rangle$ ($\left\langle E^1_d\right\rangle$) for the $0^{th}$ ($1^{st}$) order mode. In \mbox{Fig. \ref{figure4} (d)} we plot $\Delta E_1$ as a function of $\Delta E_0$ for selected disorder induced states distributed along the waveguide. We observe a strong correlation which is reflected by the linear fit and the Pearson-product-moment correlation coefficient is $0.86$. However, the slope of $1.44$ indicates that the fabrication induced disorder has a larger impact on the $1^{st}$ order mode than on the fundamental mode.


\section{\label{sec:section4}Conclusions}

In summary, we explored the radiative coupling of InGaAs QDs to the modified photonic environment of $0^{th}$ and $1^{st}$ order guided modes in PhC linear waveguides.  The modified density of states in the PhC waveguide was found to have a strong influence on the directionality and rate of spontaneous emission. Average Purcell-factors up to $\sim 2$, $\beta_{\Gamma}$-factors $\sim 90 \%$ and group indices up to $n_{g} \sim 70$ were observed at specific locations along the waveguide axis. Moreover, the impact of disorder on the slow light mode was evidenced by the observation of localised cavity modes randomly positioned along the WG axis with a frequency close to the slow light mode.  Here, disorder induced localisation resulted in high-Q modes that fluctuate with an energy bandwidth of $\delta E_{d} = \unit{6}{\milli\electronvolt}$ around the slow light bandedge with $Q$-factors up to $\sim 8000$. Nevertheless, the most efficient radiative loss channel for the localised cavity modes was found to be radiation into propagating waveguide modes. Our results demonstrate that slow light phenomena can be exploited for future integrated quantum circuits but that design tolerances must be able to account for the disorder induced localisation and, thereby, mechanisms to tune the local electronic and photonic properties would still be needed.


\begin{acknowledgments}
Many thanks to V. Savona and M. Minkov for useful discussions and we gratefully acknowledge financial support from the DFG via SFB-$631$ and the German excellence initiative via the Nanosystems Initiative Munich, the BMBF via project 16KIS0110, part of the Q.com-Halbleiter consortium and BaCaTeC via the project Integrated Quantum Optics. T.R. additionally acknowledges support of the TUM-GS.
\end{acknowledgments}


\bibliography{Bibliography}

\begin{thebibliography}{37}%
\makeatletter
\providecommand \@ifxundefined [1]{%
 \@ifx{#1\undefined}
}%
\providecommand \@ifnum [1]{%
 \ifnum #1\expandafter \@firstoftwo
 \else \expandafter \@secondoftwo
 \fi
}%
\providecommand \@ifx [1]{%
 \ifx #1\expandafter \@firstoftwo
 \else \expandafter \@secondoftwo
 \fi
}%
\providecommand \natexlab [1]{#1}%
\providecommand \enquote  [1]{``#1''}%
\providecommand \bibnamefont  [1]{#1}%
\providecommand \bibfnamefont [1]{#1}%
\providecommand \citenamefont [1]{#1}%
\providecommand \href@noop [0]{\@secondoftwo}%
\providecommand \href [0]{\begingroup \@sanitize@url \@href}%
\providecommand \@href[1]{\@@startlink{#1}\@@href}%
\providecommand \@@href[1]{\endgroup#1\@@endlink}%
\providecommand \@sanitize@url [0]{\catcode `\\12\catcode `\$12\catcode
  `\&12\catcode `\#12\catcode `\^12\catcode `\_12\catcode `\%12\relax}%
\providecommand \@@startlink[1]{}%
\providecommand \@@endlink[0]{}%
\providecommand \url  [0]{\begingroup\@sanitize@url \@url }%
\providecommand \@url [1]{\endgroup\@href {#1}{\urlprefix }}%
\providecommand \urlprefix  [0]{URL }%
\providecommand \Eprint [0]{\href }%
\providecommand \doibase [0]{http://dx.doi.org/}%
\providecommand \selectlanguage [0]{\@gobble}%
\providecommand \bibinfo  [0]{\@secondoftwo}%
\providecommand \bibfield  [0]{\@secondoftwo}%
\providecommand \translation [1]{[#1]}%
\providecommand \BibitemOpen [0]{}%
\providecommand \bibitemStop [0]{}%
\providecommand \bibitemNoStop [0]{.\EOS\space}%
\providecommand \EOS [0]{\spacefactor3000\relax}%
\providecommand \BibitemShut  [1]{\csname bibitem#1\endcsname}%
\let\auto@bib@innerbib\@empty
\bibitem [{\citenamefont {Knill}\ \emph {et~al.}(2001)\citenamefont {Knill},
  \citenamefont {Laflamme},\ and\ \citenamefont {Milburn}}]{Knill2001}%
  \BibitemOpen
  \bibfield  {author} {\bibinfo {author} {\bibfnamefont {E.}~\bibnamefont
  {Knill}}, \bibinfo {author} {\bibfnamefont {R.}~\bibnamefont {Laflamme}}, \
  and\ \bibinfo {author} {\bibfnamefont {G.~J.}\ \bibnamefont {Milburn}},\
  }\href {\doibase 10.1038/35051009} {\bibfield  {journal} {\bibinfo  {journal}
  {Nature}\ }\textbf {\bibinfo {volume} {409}},\ \bibinfo {pages} {46}
  (\bibinfo {year} {2001})}\BibitemShut {NoStop}%
\bibitem [{\citenamefont {Claudon}\ \emph {et~al.}(2010)\citenamefont
  {Claudon}, \citenamefont {Bleuse}, \citenamefont {Malik}, \citenamefont
  {Bazin}, \citenamefont {Jaffrennou}, \citenamefont {Gregersen}, \citenamefont
  {Sauvan}, \citenamefont {Lalanne},\ and\ \citenamefont
  {G{\'e}rard}}]{Claudon2010}%
  \BibitemOpen
  \bibfield  {author} {\bibinfo {author} {\bibfnamefont {J.}~\bibnamefont
  {Claudon}}, \bibinfo {author} {\bibfnamefont {J.}~\bibnamefont {Bleuse}},
  \bibinfo {author} {\bibfnamefont {N.~S.}\ \bibnamefont {Malik}}, \bibinfo
  {author} {\bibfnamefont {M.}~\bibnamefont {Bazin}}, \bibinfo {author}
  {\bibfnamefont {P.}~\bibnamefont {Jaffrennou}}, \bibinfo {author}
  {\bibfnamefont {N.}~\bibnamefont {Gregersen}}, \bibinfo {author}
  {\bibfnamefont {C.}~\bibnamefont {Sauvan}}, \bibinfo {author} {\bibfnamefont
  {P.}~\bibnamefont {Lalanne}}, \ and\ \bibinfo {author} {\bibfnamefont
  {J.-M.}\ \bibnamefont {G{\'e}rard}},\ }\href {\doibase
  10.1038/nphoton.2009.287} {\bibfield  {journal} {\bibinfo  {journal} {Nature
  Photonics}\ }\textbf {\bibinfo {volume} {4}},\ \bibinfo {pages} {174}
  (\bibinfo {year} {2010})}\BibitemShut {NoStop}%
\bibitem [{\citenamefont {Reimer}\ \emph {et~al.}(2012)\citenamefont {Reimer},
  \citenamefont {Bulgarini}, \citenamefont {Akopian}, \citenamefont {Hocevar},
  \citenamefont {Bavinck}, \citenamefont {Verheijen}, \citenamefont {Bakkers},
  \citenamefont {Kouwenhoven},\ and\ \citenamefont {Zwiller}}]{Reimer2012}%
  \BibitemOpen
  \bibfield  {author} {\bibinfo {author} {\bibfnamefont {M.~E.}\ \bibnamefont
  {Reimer}}, \bibinfo {author} {\bibfnamefont {G.}~\bibnamefont {Bulgarini}},
  \bibinfo {author} {\bibfnamefont {N.}~\bibnamefont {Akopian}}, \bibinfo
  {author} {\bibfnamefont {M.}~\bibnamefont {Hocevar}}, \bibinfo {author}
  {\bibfnamefont {M.~B.}\ \bibnamefont {Bavinck}}, \bibinfo {author}
  {\bibfnamefont {M.~A.}\ \bibnamefont {Verheijen}}, \bibinfo {author}
  {\bibfnamefont {E.~P.}\ \bibnamefont {Bakkers}}, \bibinfo {author}
  {\bibfnamefont {L.~P.}\ \bibnamefont {Kouwenhoven}}, \ and\ \bibinfo {author}
  {\bibfnamefont {V.}~\bibnamefont {Zwiller}},\ }\href {\doibase
  10.1038/ncomms1746} {\bibfield  {journal} {\bibinfo  {journal} {Nature
  communications}\ }\textbf {\bibinfo {volume} {3}},\ \bibinfo {pages} {737}
  (\bibinfo {year} {2012})}\BibitemShut {NoStop}%
\bibitem [{\citenamefont {O'Brien}\ \emph {et~al.}(2009)\citenamefont
  {O'Brien}, \citenamefont {Furusawa},\ and\ \citenamefont
  {Vu{\v{c}}kovi{\'c}}}]{O-Brien2009Photonic}%
  \BibitemOpen
  \bibfield  {author} {\bibinfo {author} {\bibfnamefont {J.~L.}\ \bibnamefont
  {O'Brien}}, \bibinfo {author} {\bibfnamefont {A.}~\bibnamefont {Furusawa}}, \
  and\ \bibinfo {author} {\bibfnamefont {J.}~\bibnamefont
  {Vu{\v{c}}kovi{\'c}}},\ }\href {\doibase 10.1038/nphoton.2009.229} {\bibfield
   {journal} {\bibinfo  {journal} {Nature Photonics}\ }\textbf {\bibinfo
  {volume} {3}},\ \bibinfo {pages} {687} (\bibinfo {year} {2009})}\BibitemShut
  {NoStop}%
\bibitem [{\citenamefont {Turchette}\ \emph {et~al.}(1995)\citenamefont
  {Turchette}, \citenamefont {Hood}, \citenamefont {Lange}, \citenamefont
  {Mabuchi},\ and\ \citenamefont {Kimble}}]{Turchette1995}%
  \BibitemOpen
  \bibfield  {author} {\bibinfo {author} {\bibfnamefont {Q.~A.}\ \bibnamefont
  {Turchette}}, \bibinfo {author} {\bibfnamefont {C.}~\bibnamefont {Hood}},
  \bibinfo {author} {\bibfnamefont {W.}~\bibnamefont {Lange}}, \bibinfo
  {author} {\bibfnamefont {H.}~\bibnamefont {Mabuchi}}, \ and\ \bibinfo
  {author} {\bibfnamefont {H.~J.}\ \bibnamefont {Kimble}},\ }\href {\doibase
  http://dx.doi.org/10.1103/PhysRevLett.75.4710} {\bibfield  {journal}
  {\bibinfo  {journal} {Physical Review Letters}\ }\textbf {\bibinfo {volume}
  {75}},\ \bibinfo {pages} {4710} (\bibinfo {year} {1995})}\BibitemShut
  {NoStop}%
\bibitem [{\citenamefont {Nogues}\ \emph {et~al.}(1999)\citenamefont {Nogues},
  \citenamefont {Rauschenbeutel}, \citenamefont {Osnaghi}, \citenamefont
  {Brune}, \citenamefont {Raimond},\ and\ \citenamefont
  {Haroche}}]{Nogues1999}%
  \BibitemOpen
  \bibfield  {author} {\bibinfo {author} {\bibfnamefont {G.}~\bibnamefont
  {Nogues}}, \bibinfo {author} {\bibfnamefont {A.}~\bibnamefont
  {Rauschenbeutel}}, \bibinfo {author} {\bibfnamefont {S.}~\bibnamefont
  {Osnaghi}}, \bibinfo {author} {\bibfnamefont {M.}~\bibnamefont {Brune}},
  \bibinfo {author} {\bibfnamefont {J.}~\bibnamefont {Raimond}}, \ and\
  \bibinfo {author} {\bibfnamefont {S.}~\bibnamefont {Haroche}},\ }\href
  {\doibase 10.1038/22275} {\bibfield  {journal} {\bibinfo  {journal} {Nature}\
  }\textbf {\bibinfo {volume} {400}},\ \bibinfo {pages} {239} (\bibinfo {year}
  {1999})}\BibitemShut {NoStop}%
\bibitem [{\citenamefont {Englund}\ \emph {et~al.}(2007)\citenamefont
  {Englund}, \citenamefont {Faraon}, \citenamefont {Fushman}, \citenamefont
  {Stoltz}, \citenamefont {Petroff},\ and\ \citenamefont
  {Vu{\v{c}}kovi{\'c}}}]{Englund2007Cav_Ref}%
  \BibitemOpen
  \bibfield  {author} {\bibinfo {author} {\bibfnamefont {D.}~\bibnamefont
  {Englund}}, \bibinfo {author} {\bibfnamefont {A.}~\bibnamefont {Faraon}},
  \bibinfo {author} {\bibfnamefont {I.}~\bibnamefont {Fushman}}, \bibinfo
  {author} {\bibfnamefont {N.}~\bibnamefont {Stoltz}}, \bibinfo {author}
  {\bibfnamefont {P.}~\bibnamefont {Petroff}}, \ and\ \bibinfo {author}
  {\bibfnamefont {J.}~\bibnamefont {Vu{\v{c}}kovi{\'c}}},\ }\href {\doibase
  10.1038/nature06234} {\bibfield  {journal} {\bibinfo  {journal} {Nature}\
  }\textbf {\bibinfo {volume} {450}},\ \bibinfo {pages} {857} (\bibinfo {year}
  {2007})}\BibitemShut {NoStop}%
\bibitem [{\citenamefont {Fushman}\ \emph {et~al.}(2008)\citenamefont
  {Fushman}, \citenamefont {Englund}, \citenamefont {Faraon}, \citenamefont
  {Stoltz}, \citenamefont {Petroff},\ and\ \citenamefont
  {Vu{\v{c}}kovi{\'c}}}]{Fushman2008}%
  \BibitemOpen
  \bibfield  {author} {\bibinfo {author} {\bibfnamefont {I.}~\bibnamefont
  {Fushman}}, \bibinfo {author} {\bibfnamefont {D.}~\bibnamefont {Englund}},
  \bibinfo {author} {\bibfnamefont {A.}~\bibnamefont {Faraon}}, \bibinfo
  {author} {\bibfnamefont {N.}~\bibnamefont {Stoltz}}, \bibinfo {author}
  {\bibfnamefont {P.}~\bibnamefont {Petroff}}, \ and\ \bibinfo {author}
  {\bibfnamefont {J.}~\bibnamefont {Vu{\v{c}}kovi{\'c}}},\ }\href {\doibase
  10.1126/science.1154643} {\bibfield  {journal} {\bibinfo  {journal}
  {Science}\ }\textbf {\bibinfo {volume} {320}},\ \bibinfo {pages} {769}
  (\bibinfo {year} {2008})}\BibitemShut {NoStop}%
\bibitem [{\citenamefont {Englund}\ \emph {et~al.}(2010)\citenamefont
  {Englund}, \citenamefont {Shields}, \citenamefont {Rivoire}, \citenamefont
  {Hatami}, \citenamefont {Vuckovic}, \citenamefont {Park},\ and\ \citenamefont
  {Lukin}}]{Englund2010}%
  \BibitemOpen
  \bibfield  {author} {\bibinfo {author} {\bibfnamefont {D.}~\bibnamefont
  {Englund}}, \bibinfo {author} {\bibfnamefont {B.}~\bibnamefont {Shields}},
  \bibinfo {author} {\bibfnamefont {K.}~\bibnamefont {Rivoire}}, \bibinfo
  {author} {\bibfnamefont {F.}~\bibnamefont {Hatami}}, \bibinfo {author}
  {\bibfnamefont {J.}~\bibnamefont {Vuckovic}}, \bibinfo {author}
  {\bibfnamefont {H.}~\bibnamefont {Park}}, \ and\ \bibinfo {author}
  {\bibfnamefont {M.~D.}\ \bibnamefont {Lukin}},\ }\href {\doibase
  10.1021/nl101662v} {\bibfield  {journal} {\bibinfo  {journal} {Nano letters}\
  }\textbf {\bibinfo {volume} {10}},\ \bibinfo {pages} {3922} (\bibinfo {year}
  {2010})}\BibitemShut {NoStop}%
\bibitem [{\citenamefont {Volz}\ \emph {et~al.}(2012)\citenamefont {Volz},
  \citenamefont {Reinhard}, \citenamefont {Winger}, \citenamefont {Badolato},
  \citenamefont {Hennessy}, \citenamefont {Hu},\ and\ \citenamefont
  {Imamo{\u{g}}lu}}]{Volz2012}%
  \BibitemOpen
  \bibfield  {author} {\bibinfo {author} {\bibfnamefont {T.}~\bibnamefont
  {Volz}}, \bibinfo {author} {\bibfnamefont {A.}~\bibnamefont {Reinhard}},
  \bibinfo {author} {\bibfnamefont {M.}~\bibnamefont {Winger}}, \bibinfo
  {author} {\bibfnamefont {A.}~\bibnamefont {Badolato}}, \bibinfo {author}
  {\bibfnamefont {K.~J.}\ \bibnamefont {Hennessy}}, \bibinfo {author}
  {\bibfnamefont {E.~L.}\ \bibnamefont {Hu}}, \ and\ \bibinfo {author}
  {\bibfnamefont {A.}~\bibnamefont {Imamo{\u{g}}lu}},\ }\href {\doibase
  10.1038/nphoton.2012.181} {\bibfield  {journal} {\bibinfo  {journal} {Nature
  Photonics}\ }\textbf {\bibinfo {volume} {6}},\ \bibinfo {pages} {605}
  (\bibinfo {year} {2012})}\BibitemShut {NoStop}%
\bibitem [{\citenamefont {Englund}\ \emph {et~al.}(2012)\citenamefont
  {Englund}, \citenamefont {Majumdar}, \citenamefont {Bajcsy}, \citenamefont
  {Faraon}, \citenamefont {Petroff},\ and\ \citenamefont
  {Vu{\v{c}}kovi{\'c}}}]{Englund2012}%
  \BibitemOpen
  \bibfield  {author} {\bibinfo {author} {\bibfnamefont {D.}~\bibnamefont
  {Englund}}, \bibinfo {author} {\bibfnamefont {A.}~\bibnamefont {Majumdar}},
  \bibinfo {author} {\bibfnamefont {M.}~\bibnamefont {Bajcsy}}, \bibinfo
  {author} {\bibfnamefont {A.}~\bibnamefont {Faraon}}, \bibinfo {author}
  {\bibfnamefont {P.}~\bibnamefont {Petroff}}, \ and\ \bibinfo {author}
  {\bibfnamefont {J.}~\bibnamefont {Vu{\v{c}}kovi{\'c}}},\ }\href
  {http://journals.aps.org/prl/abstract/10.1103/PhysRevLett.108.093604}
  {\bibfield  {journal} {\bibinfo  {journal} {Physical review letters}\
  }\textbf {\bibinfo {volume} {108}},\ \bibinfo {pages} {093604} (\bibinfo
  {year} {2012})}\BibitemShut {NoStop}%
\bibitem [{\citenamefont {Hwang}\ \emph {et~al.}(2009)\citenamefont {Hwang},
  \citenamefont {Pototschnig}, \citenamefont {Lettow}, \citenamefont {Zumofen},
  \citenamefont {Renn}, \citenamefont {G{\"o}tzinger},\ and\ \citenamefont
  {Sandoghdar}}]{Hwang2009}%
  \BibitemOpen
  \bibfield  {author} {\bibinfo {author} {\bibfnamefont {J.}~\bibnamefont
  {Hwang}}, \bibinfo {author} {\bibfnamefont {M.}~\bibnamefont {Pototschnig}},
  \bibinfo {author} {\bibfnamefont {R.}~\bibnamefont {Lettow}}, \bibinfo
  {author} {\bibfnamefont {G.}~\bibnamefont {Zumofen}}, \bibinfo {author}
  {\bibfnamefont {A.}~\bibnamefont {Renn}}, \bibinfo {author} {\bibfnamefont
  {S.}~\bibnamefont {G{\"o}tzinger}}, \ and\ \bibinfo {author} {\bibfnamefont
  {V.}~\bibnamefont {Sandoghdar}},\ }\href {\doibase 10.1038/nature08134}
  {\bibfield  {journal} {\bibinfo  {journal} {Nature}\ }\textbf {\bibinfo
  {volume} {460}},\ \bibinfo {pages} {76} (\bibinfo {year} {2009})}\BibitemShut
  {NoStop}%
\bibitem [{\citenamefont {Honjo}\ \emph {et~al.}(2004)\citenamefont {Honjo},
  \citenamefont {Inoue},\ and\ \citenamefont {Takahashi}}]{Honjo2004}%
  \BibitemOpen
  \bibfield  {author} {\bibinfo {author} {\bibfnamefont {T.}~\bibnamefont
  {Honjo}}, \bibinfo {author} {\bibfnamefont {K.}~\bibnamefont {Inoue}}, \ and\
  \bibinfo {author} {\bibfnamefont {H.}~\bibnamefont {Takahashi}},\ }\href
  {\doibase http://dx.doi.org/10.1364/OL.29.002797} {\bibfield  {journal}
  {\bibinfo  {journal} {Optics letters}\ }\textbf {\bibinfo {volume} {29}},\
  \bibinfo {pages} {2797} (\bibinfo {year} {2004})}\BibitemShut {NoStop}%
\bibitem [{\citenamefont {Takesue}\ and\ \citenamefont
  {Inoue}(2005)}]{Takesue2005}%
  \BibitemOpen
  \bibfield  {author} {\bibinfo {author} {\bibfnamefont {H.}~\bibnamefont
  {Takesue}}\ and\ \bibinfo {author} {\bibfnamefont {K.}~\bibnamefont
  {Inoue}},\ }\href {\doibase http://dx.doi.org/10.1364/OPEX.13.007832}
  {\bibfield  {journal} {\bibinfo  {journal} {Optics express}\ }\textbf
  {\bibinfo {volume} {13}},\ \bibinfo {pages} {7832} (\bibinfo {year}
  {2005})}\BibitemShut {NoStop}%
\bibitem [{\citenamefont {Mu{\~n}oz}\ \emph {et~al.}(2014)\citenamefont
  {Mu{\~n}oz}, \citenamefont {del Valle}, \citenamefont {Tudela}, \citenamefont
  {M{\"u}ller}, \citenamefont {Lichtmannecker}, \citenamefont {Kaniber},
  \citenamefont {Tejedor}, \citenamefont {Finley},\ and\ \citenamefont
  {Laussy}}]{Munoz2014}%
  \BibitemOpen
  \bibfield  {author} {\bibinfo {author} {\bibfnamefont {C.~S.}\ \bibnamefont
  {Mu{\~n}oz}}, \bibinfo {author} {\bibfnamefont {E.}~\bibnamefont {del
  Valle}}, \bibinfo {author} {\bibfnamefont {A.~G.}\ \bibnamefont {Tudela}},
  \bibinfo {author} {\bibfnamefont {K.}~\bibnamefont {M{\"u}ller}}, \bibinfo
  {author} {\bibfnamefont {S.}~\bibnamefont {Lichtmannecker}}, \bibinfo
  {author} {\bibfnamefont {M.}~\bibnamefont {Kaniber}}, \bibinfo {author}
  {\bibfnamefont {C.}~\bibnamefont {Tejedor}}, \bibinfo {author} {\bibfnamefont
  {J.}~\bibnamefont {Finley}}, \ and\ \bibinfo {author} {\bibfnamefont
  {F.}~\bibnamefont {Laussy}},\ }\href {\doibase doi:10.1038/nphoton.2014.114}
  {\bibfield  {journal} {\bibinfo  {journal} {Nature Photonics}\ } (\bibinfo
  {year} {2014}),\ doi:10.1038/nphoton.2014.114}\BibitemShut {NoStop}%
\bibitem [{\citenamefont {Yao}\ and\ \citenamefont
  {Hughes}(2009)}]{Yao2009Cav-WG}%
  \BibitemOpen
  \bibfield  {author} {\bibinfo {author} {\bibfnamefont {P.}~\bibnamefont
  {Yao}}\ and\ \bibinfo {author} {\bibfnamefont {S.}~\bibnamefont {Hughes}},\
  }\href {\doibase 10.1103/PhysRevB.80.165128} {\bibfield  {journal} {\bibinfo
  {journal} {Physical Review B}\ }\textbf {\bibinfo {volume} {80}},\ \bibinfo
  {pages} {165128} (\bibinfo {year} {2009})}\BibitemShut {NoStop}%
\bibitem [{\citenamefont {Rao}\ and\ \citenamefont
  {Hughes}(2007{\natexlab{a}})}]{Rao2007Finite}%
  \BibitemOpen
  \bibfield  {author} {\bibinfo {author} {\bibfnamefont {V.~M.}\ \bibnamefont
  {Rao}}\ and\ \bibinfo {author} {\bibfnamefont {S.}~\bibnamefont {Hughes}},\
  }\href {\doibase http://dx.doi.org/10.1103/PhysRevLett.99.193901} {\bibfield
  {journal} {\bibinfo  {journal} {Physical review letters}\ }\textbf {\bibinfo
  {volume} {99}},\ \bibinfo {pages} {193901} (\bibinfo {year}
  {2007}{\natexlab{a}})}\BibitemShut {NoStop}%
\bibitem [{\citenamefont {Hughes}(2004)}]{Hughes2004}%
  \BibitemOpen
  \bibfield  {author} {\bibinfo {author} {\bibfnamefont {S.}~\bibnamefont
  {Hughes}},\ }\href {\doibase 10.1364/OL.29.002659} {\bibfield  {journal}
  {\bibinfo  {journal} {Optics letters}\ }\textbf {\bibinfo {volume} {29}},\
  \bibinfo {pages} {2659} (\bibinfo {year} {2004})}\BibitemShut {NoStop}%
\bibitem [{\citenamefont {Viasnoff-Schwoob}\ \emph {et~al.}(2005)\citenamefont
  {Viasnoff-Schwoob}, \citenamefont {Weisbuch}, \citenamefont {Benisty},
  \citenamefont {Olivier}, \citenamefont {Varoutsis}, \citenamefont
  {Robert-Philip}, \citenamefont {Houdr{\'e}},\ and\ \citenamefont
  {Smith}}]{Viasnoff2005}%
  \BibitemOpen
  \bibfield  {author} {\bibinfo {author} {\bibfnamefont {E.}~\bibnamefont
  {Viasnoff-Schwoob}}, \bibinfo {author} {\bibfnamefont {C.}~\bibnamefont
  {Weisbuch}}, \bibinfo {author} {\bibfnamefont {H.}~\bibnamefont {Benisty}},
  \bibinfo {author} {\bibfnamefont {S.}~\bibnamefont {Olivier}}, \bibinfo
  {author} {\bibfnamefont {S.}~\bibnamefont {Varoutsis}}, \bibinfo {author}
  {\bibfnamefont {I.}~\bibnamefont {Robert-Philip}}, \bibinfo {author}
  {\bibfnamefont {R.}~\bibnamefont {Houdr{\'e}}}, \ and\ \bibinfo {author}
  {\bibfnamefont {C.}~\bibnamefont {Smith}},\ }\href {\doibase
  10.1103/PhysRevLett.95.183901} {\bibfield  {journal} {\bibinfo  {journal}
  {Physical review letters}\ }\textbf {\bibinfo {volume} {95}},\ \bibinfo
  {pages} {183901} (\bibinfo {year} {2005})}\BibitemShut {NoStop}%
\bibitem [{\citenamefont {Lecamp}\ \emph {et~al.}(2007)\citenamefont {Lecamp},
  \citenamefont {Lalanne},\ and\ \citenamefont {Hugonin}}]{Lecamp2007}%
  \BibitemOpen
  \bibfield  {author} {\bibinfo {author} {\bibfnamefont {G.}~\bibnamefont
  {Lecamp}}, \bibinfo {author} {\bibfnamefont {P.}~\bibnamefont {Lalanne}}, \
  and\ \bibinfo {author} {\bibfnamefont {J.}~\bibnamefont {Hugonin}},\ }\href
  {\doibase 10.1103/PhysRevLett.99.023902} {\bibfield  {journal} {\bibinfo
  {journal} {Physical review letters}\ }\textbf {\bibinfo {volume} {99}},\
  \bibinfo {pages} {023902} (\bibinfo {year} {2007})}\BibitemShut {NoStop}%
\bibitem [{\citenamefont {Vlasov}\ \emph {et~al.}(2005)\citenamefont {Vlasov},
  \citenamefont {O'Boyle}, \citenamefont {Hamann},\ and\ \citenamefont
  {McNab}}]{Vlasov2005}%
  \BibitemOpen
  \bibfield  {author} {\bibinfo {author} {\bibfnamefont {Y.~A.}\ \bibnamefont
  {Vlasov}}, \bibinfo {author} {\bibfnamefont {M.}~\bibnamefont {O'Boyle}},
  \bibinfo {author} {\bibfnamefont {H.~F.}\ \bibnamefont {Hamann}}, \ and\
  \bibinfo {author} {\bibfnamefont {S.~J.}\ \bibnamefont {McNab}},\ }\href
  {\doibase 10.1038/nature04210} {\bibfield  {journal} {\bibinfo  {journal}
  {Nature}\ }\textbf {\bibinfo {volume} {438}},\ \bibinfo {pages} {65}
  (\bibinfo {year} {2005})}\BibitemShut {NoStop}%
\bibitem [{\citenamefont {Lund-Hansen}\ \emph {et~al.}(2008)\citenamefont
  {Lund-Hansen}, \citenamefont {Stobbe}, \citenamefont {Julsgaard},
  \citenamefont {Thyrrestrup}, \citenamefont {S{\"u}nner}, \citenamefont
  {Kamp}, \citenamefont {Forchel},\ and\ \citenamefont
  {Lodahl}}]{Lund-Hansen2008}%
  \BibitemOpen
  \bibfield  {author} {\bibinfo {author} {\bibfnamefont {T.}~\bibnamefont
  {Lund-Hansen}}, \bibinfo {author} {\bibfnamefont {S.}~\bibnamefont {Stobbe}},
  \bibinfo {author} {\bibfnamefont {B.}~\bibnamefont {Julsgaard}}, \bibinfo
  {author} {\bibfnamefont {H.}~\bibnamefont {Thyrrestrup}}, \bibinfo {author}
  {\bibfnamefont {T.}~\bibnamefont {S{\"u}nner}}, \bibinfo {author}
  {\bibfnamefont {M.}~\bibnamefont {Kamp}}, \bibinfo {author} {\bibfnamefont
  {A.}~\bibnamefont {Forchel}}, \ and\ \bibinfo {author} {\bibfnamefont
  {P.}~\bibnamefont {Lodahl}},\ }\href {\doibase
  http://dx.doi.org/10.1103/PhysRevLett.101.113903} {\bibfield  {journal}
  {\bibinfo  {journal} {Physical review letters}\ }\textbf {\bibinfo {volume}
  {101}},\ \bibinfo {pages} {113903} (\bibinfo {year} {2008})}\BibitemShut
  {NoStop}%
\bibitem [{\citenamefont {Thyrrestrup}\ \emph {et~al.}(2010)\citenamefont
  {Thyrrestrup}, \citenamefont {Sapienza},\ and\ \citenamefont
  {Lodahl}}]{Thyrrestrup2010}%
  \BibitemOpen
  \bibfield  {author} {\bibinfo {author} {\bibfnamefont {H.}~\bibnamefont
  {Thyrrestrup}}, \bibinfo {author} {\bibfnamefont {L.}~\bibnamefont
  {Sapienza}}, \ and\ \bibinfo {author} {\bibfnamefont {P.}~\bibnamefont
  {Lodahl}},\ }\href {\doibase 10.1063/1.3446873} {\bibfield  {journal}
  {\bibinfo  {journal} {Applied Physics Letters}\ }\textbf {\bibinfo {volume}
  {96}},\ \bibinfo {pages} {231106} (\bibinfo {year} {2010})}\BibitemShut
  {NoStop}%
\bibitem [{\citenamefont {Dewhurst}\ \emph {et~al.}(2010)\citenamefont
  {Dewhurst}, \citenamefont {Granados}, \citenamefont {Ellis}, \citenamefont
  {Bennett}, \citenamefont {Patel}, \citenamefont {Farrer}, \citenamefont
  {Anderson}, \citenamefont {Jones}, \citenamefont {Ritchie},\ and\
  \citenamefont {Shields}}]{Dewhurst2010}%
  \BibitemOpen
  \bibfield  {author} {\bibinfo {author} {\bibfnamefont {S.}~\bibnamefont
  {Dewhurst}}, \bibinfo {author} {\bibfnamefont {D.}~\bibnamefont {Granados}},
  \bibinfo {author} {\bibfnamefont {D.}~\bibnamefont {Ellis}}, \bibinfo
  {author} {\bibfnamefont {A.}~\bibnamefont {Bennett}}, \bibinfo {author}
  {\bibfnamefont {R.}~\bibnamefont {Patel}}, \bibinfo {author} {\bibfnamefont
  {I.}~\bibnamefont {Farrer}}, \bibinfo {author} {\bibfnamefont
  {D.}~\bibnamefont {Anderson}}, \bibinfo {author} {\bibfnamefont
  {G.}~\bibnamefont {Jones}}, \bibinfo {author} {\bibfnamefont
  {D.}~\bibnamefont {Ritchie}}, \ and\ \bibinfo {author} {\bibfnamefont
  {A.}~\bibnamefont {Shields}},\ }\href {\doibase
  http://dx.doi.org/10.1063/1.3294298} {\bibfield  {journal} {\bibinfo
  {journal} {Applied Physics Letters}\ }\textbf {\bibinfo {volume} {96}},\
  \bibinfo {pages} {031109} (\bibinfo {year} {2010})}\BibitemShut {NoStop}%
\bibitem [{\citenamefont {Ba~Hoang}\ \emph {et~al.}(2012)\citenamefont
  {Ba~Hoang}, \citenamefont {Beetz}, \citenamefont {Midolo}, \citenamefont
  {Skacel}, \citenamefont {Lermer}, \citenamefont {Kamp}, \citenamefont
  {Hofling}, \citenamefont {Balet}, \citenamefont {Chauvin},\ and\
  \citenamefont {Fiore}}]{Hoang2012}%
  \BibitemOpen
  \bibfield  {author} {\bibinfo {author} {\bibfnamefont {T.}~\bibnamefont
  {Ba~Hoang}}, \bibinfo {author} {\bibfnamefont {J.}~\bibnamefont {Beetz}},
  \bibinfo {author} {\bibfnamefont {L.}~\bibnamefont {Midolo}}, \bibinfo
  {author} {\bibfnamefont {M.}~\bibnamefont {Skacel}}, \bibinfo {author}
  {\bibfnamefont {M.}~\bibnamefont {Lermer}}, \bibinfo {author} {\bibfnamefont
  {M.}~\bibnamefont {Kamp}}, \bibinfo {author} {\bibfnamefont {S.}~\bibnamefont
  {Hofling}}, \bibinfo {author} {\bibfnamefont {L.}~\bibnamefont {Balet}},
  \bibinfo {author} {\bibfnamefont {N.}~\bibnamefont {Chauvin}}, \ and\
  \bibinfo {author} {\bibfnamefont {A.}~\bibnamefont {Fiore}},\ }\href
  {\doibase 10.1063/1.3683541} {\bibfield  {journal} {\bibinfo  {journal}
  {Applied Physics Letters}\ }\textbf {\bibinfo {volume} {100}},\ \bibinfo
  {pages} {061122} (\bibinfo {year} {2012})}\BibitemShut {NoStop}%
\bibitem [{\citenamefont {Gao}\ \emph {et~al.}(2008)\citenamefont {Gao},
  \citenamefont {Sun},\ and\ \citenamefont {Wong}}]{Gao2008}%
  \BibitemOpen
  \bibfield  {author} {\bibinfo {author} {\bibfnamefont {J.}~\bibnamefont
  {Gao}}, \bibinfo {author} {\bibfnamefont {F.}~\bibnamefont {Sun}}, \ and\
  \bibinfo {author} {\bibfnamefont {C.~W.}\ \bibnamefont {Wong}},\ }\href
  {\doibase http://dx.doi.org/10.1063/1.2999588} {\bibfield  {journal}
  {\bibinfo  {journal} {Applied Physics Letters}\ }\textbf {\bibinfo {volume}
  {93}},\ \bibinfo {pages} {151108} (\bibinfo {year} {2008})}\BibitemShut
  {NoStop}%
\bibitem [{\citenamefont {Gao}\ \emph {et~al.}(2013)\citenamefont {Gao},
  \citenamefont {Combrie}, \citenamefont {Liang}, \citenamefont
  {Schmitteckert}, \citenamefont {Lehoucq}, \citenamefont {Xavier},
  \citenamefont {Xu}, \citenamefont {Busch}, \citenamefont {Huffaker},
  \citenamefont {De~Rossi} \emph {et~al.}}]{Gao2013}%
  \BibitemOpen
  \bibfield  {author} {\bibinfo {author} {\bibfnamefont {J.}~\bibnamefont
  {Gao}}, \bibinfo {author} {\bibfnamefont {S.}~\bibnamefont {Combrie}},
  \bibinfo {author} {\bibfnamefont {B.}~\bibnamefont {Liang}}, \bibinfo
  {author} {\bibfnamefont {P.}~\bibnamefont {Schmitteckert}}, \bibinfo {author}
  {\bibfnamefont {G.}~\bibnamefont {Lehoucq}}, \bibinfo {author} {\bibfnamefont
  {S.}~\bibnamefont {Xavier}}, \bibinfo {author} {\bibfnamefont
  {X.}~\bibnamefont {Xu}}, \bibinfo {author} {\bibfnamefont {K.}~\bibnamefont
  {Busch}}, \bibinfo {author} {\bibfnamefont {D.~L.}\ \bibnamefont {Huffaker}},
  \bibinfo {author} {\bibfnamefont {A.}~\bibnamefont {De~Rossi}},  \emph
  {et~al.},\ }\href {\doibase doi:10.1038/srep01994} {\bibfield  {journal}
  {\bibinfo  {journal} {Scientific reports}\ }\textbf {\bibinfo {volume} {3}}
  (\bibinfo {year} {2013}),\ doi:10.1038/srep01994}\BibitemShut {NoStop}%
\bibitem [{\citenamefont {Savona}(2011)}]{Savona2011}%
  \BibitemOpen
  \bibfield  {author} {\bibinfo {author} {\bibfnamefont {V.}~\bibnamefont
  {Savona}},\ }\href {\doibase 10.1103/PhysRevB.83.085301} {\bibfield
  {journal} {\bibinfo  {journal} {Physical Review B}\ }\textbf {\bibinfo
  {volume} {83}},\ \bibinfo {pages} {085301} (\bibinfo {year}
  {2011})}\BibitemShut {NoStop}%
\bibitem [{\citenamefont {Huisman}\ \emph {et~al.}(2012)\citenamefont
  {Huisman}, \citenamefont {Ctistis}, \citenamefont {Stobbe}, \citenamefont
  {Mosk}, \citenamefont {Herek}, \citenamefont {Lagendijk}, \citenamefont
  {Lodahl}, \citenamefont {Vos},\ and\ \citenamefont {Pinkse}}]{Huisman2012}%
  \BibitemOpen
  \bibfield  {author} {\bibinfo {author} {\bibfnamefont {S.}~\bibnamefont
  {Huisman}}, \bibinfo {author} {\bibfnamefont {G.}~\bibnamefont {Ctistis}},
  \bibinfo {author} {\bibfnamefont {S.}~\bibnamefont {Stobbe}}, \bibinfo
  {author} {\bibfnamefont {A.}~\bibnamefont {Mosk}}, \bibinfo {author}
  {\bibfnamefont {J.}~\bibnamefont {Herek}}, \bibinfo {author} {\bibfnamefont
  {A.}~\bibnamefont {Lagendijk}}, \bibinfo {author} {\bibfnamefont
  {P.}~\bibnamefont {Lodahl}}, \bibinfo {author} {\bibfnamefont
  {W.}~\bibnamefont {Vos}}, \ and\ \bibinfo {author} {\bibfnamefont
  {P.}~\bibnamefont {Pinkse}},\ }\href {\doibase
  http://dx.doi.org/10.1103/PhysRevB.86.155154} {\bibfield  {journal} {\bibinfo
   {journal} {Physical Review B}\ }\textbf {\bibinfo {volume} {86}},\ \bibinfo
  {pages} {155154} (\bibinfo {year} {2012})}\BibitemShut {NoStop}%
\bibitem [{\citenamefont {Lagendijk}\ \emph {et~al.}(2009)\citenamefont
  {Lagendijk}, \citenamefont {van Tiggelen},\ and\ \citenamefont
  {Wiersma}}]{Lagendijk2009}%
  \BibitemOpen
  \bibfield  {author} {\bibinfo {author} {\bibfnamefont {A.}~\bibnamefont
  {Lagendijk}}, \bibinfo {author} {\bibfnamefont {B.}~\bibnamefont {van
  Tiggelen}}, \ and\ \bibinfo {author} {\bibfnamefont {D.~S.}\ \bibnamefont
  {Wiersma}},\ }\href {\doibase http://dx.doi.org/10.1063/1.3206091} {\bibfield
   {journal} {\bibinfo  {journal} {Phys. Today}\ }\textbf {\bibinfo {volume}
  {62}},\ \bibinfo {pages} {24} (\bibinfo {year} {2009})}\BibitemShut {NoStop}%
\bibitem [{\citenamefont {RSoft}()}]{RSoft2014}%
  \BibitemOpen
  \bibfield  {author} {\bibinfo {author} {\bibnamefont {RSoft}},\ }\href
  {http://optics.synopsys.com/rsoft/} {}\bibinfo {howpublished}
  {\url{http://optics.synopsys.com/rsoft/}},\ \bibinfo {note} {[Online
  accessed: 24-July-2014]}\BibitemShut {NoStop}%
\bibitem [{\citenamefont {Johnson}\ \emph {et~al.}(2000)\citenamefont
  {Johnson}, \citenamefont {Villeneuve}, \citenamefont {Fan},\ and\
  \citenamefont {Joannopoulos}}]{Johnson2000}%
  \BibitemOpen
  \bibfield  {author} {\bibinfo {author} {\bibfnamefont {S.~G.}\ \bibnamefont
  {Johnson}}, \bibinfo {author} {\bibfnamefont {P.~R.}\ \bibnamefont
  {Villeneuve}}, \bibinfo {author} {\bibfnamefont {S.}~\bibnamefont {Fan}}, \
  and\ \bibinfo {author} {\bibfnamefont {J.}~\bibnamefont {Joannopoulos}},\
  }\href {\doibase 10.1103/PhysRevB.62.8212} {\bibfield  {journal} {\bibinfo
  {journal} {Physical Review B}\ }\textbf {\bibinfo {volume} {62}},\ \bibinfo
  {pages} {8212} (\bibinfo {year} {2000})}\BibitemShut {NoStop}%
\bibitem [{\citenamefont {Dorfner}\ \emph {et~al.}(2008)\citenamefont
  {Dorfner}, \citenamefont {Hurlimann}, \citenamefont {Zabel}, \citenamefont
  {Frandsen}, \citenamefont {Abstreiter},\ and\ \citenamefont
  {Finley}}]{Dorfner2008}%
  \BibitemOpen
  \bibfield  {author} {\bibinfo {author} {\bibfnamefont {D.}~\bibnamefont
  {Dorfner}}, \bibinfo {author} {\bibfnamefont {T.}~\bibnamefont {Hurlimann}},
  \bibinfo {author} {\bibfnamefont {T.}~\bibnamefont {Zabel}}, \bibinfo
  {author} {\bibfnamefont {L.~H.}\ \bibnamefont {Frandsen}}, \bibinfo {author}
  {\bibfnamefont {G.}~\bibnamefont {Abstreiter}}, \ and\ \bibinfo {author}
  {\bibfnamefont {J.}~\bibnamefont {Finley}},\ }\href {\doibase doi:
  10.1063/1.3009203} {\bibfield  {journal} {\bibinfo  {journal} {Applied
  Physics Letters}\ }\textbf {\bibinfo {volume} {93}},\ \bibinfo {pages}
  {181103} (\bibinfo {year} {2008})}\BibitemShut {NoStop}%
\bibitem [{\citenamefont {Laucht}\ \emph
  {et~al.}(2012{\natexlab{a}})\citenamefont {Laucht}, \citenamefont
  {G{\"u}nthner}, \citenamefont {P{\"u}tz}, \citenamefont {Saive},
  \citenamefont {Frederick}, \citenamefont {Hauke}, \citenamefont {Bichler},
  \citenamefont {Amann}, \citenamefont {Holleitner}, \citenamefont {Kaniber},\
  and\ \citenamefont {Finley}}]{Laucht2012Ensemble}%
  \BibitemOpen
  \bibfield  {author} {\bibinfo {author} {\bibfnamefont {A.}~\bibnamefont
  {Laucht}}, \bibinfo {author} {\bibfnamefont {T.}~\bibnamefont
  {G{\"u}nthner}}, \bibinfo {author} {\bibfnamefont {S.}~\bibnamefont
  {P{\"u}tz}}, \bibinfo {author} {\bibfnamefont {R.}~\bibnamefont {Saive}},
  \bibinfo {author} {\bibfnamefont {S.}~\bibnamefont {Frederick}}, \bibinfo
  {author} {\bibfnamefont {N.}~\bibnamefont {Hauke}}, \bibinfo {author}
  {\bibfnamefont {M.}~\bibnamefont {Bichler}}, \bibinfo {author} {\bibfnamefont
  {M.-C.}\ \bibnamefont {Amann}}, \bibinfo {author} {\bibfnamefont
  {A.}~\bibnamefont {Holleitner}}, \bibinfo {author} {\bibfnamefont
  {M.}~\bibnamefont {Kaniber}}, \ and\ \bibinfo {author} {\bibfnamefont
  {J.}~\bibnamefont {Finley}},\ }\href {\doibase 10.1063/1.4764923} {\bibfield
  {journal} {\bibinfo  {journal} {Journal of Applied Physics}\ }\textbf
  {\bibinfo {volume} {112}},\ \bibinfo {pages} {093520} (\bibinfo {year}
  {2012}{\natexlab{a}})}\BibitemShut {NoStop}%
\bibitem [{\citenamefont {Hohenester}\ \emph {et~al.}(2009)\citenamefont
  {Hohenester}, \citenamefont {Laucht}, \citenamefont {Kaniber}, \citenamefont
  {Hauke}, \citenamefont {Neumann}, \citenamefont {Mohtashami}, \citenamefont
  {Seliger}, \citenamefont {Bichler},\ and\ \citenamefont
  {Finley}}]{Hohenester2009}%
  \BibitemOpen
  \bibfield  {author} {\bibinfo {author} {\bibfnamefont {U.}~\bibnamefont
  {Hohenester}}, \bibinfo {author} {\bibfnamefont {A.}~\bibnamefont {Laucht}},
  \bibinfo {author} {\bibfnamefont {M.}~\bibnamefont {Kaniber}}, \bibinfo
  {author} {\bibfnamefont {N.}~\bibnamefont {Hauke}}, \bibinfo {author}
  {\bibfnamefont {A.}~\bibnamefont {Neumann}}, \bibinfo {author} {\bibfnamefont
  {A.}~\bibnamefont {Mohtashami}}, \bibinfo {author} {\bibfnamefont
  {M.}~\bibnamefont {Seliger}}, \bibinfo {author} {\bibfnamefont
  {M.}~\bibnamefont {Bichler}}, \ and\ \bibinfo {author} {\bibfnamefont
  {J.~J.}\ \bibnamefont {Finley}},\ }\href {\doibase
  10.1103/PhysRevB.80.201311} {\bibfield  {journal} {\bibinfo  {journal}
  {Physical Review B}\ }\textbf {\bibinfo {volume} {80}},\ \bibinfo {pages}
  {201311} (\bibinfo {year} {2009})}\BibitemShut {NoStop}%
\bibitem [{\citenamefont {Laucht}\ \emph
  {et~al.}(2012{\natexlab{b}})\citenamefont {Laucht}, \citenamefont {P{\"u}tz},
  \citenamefont {G{\"u}nthner}, \citenamefont {Hauke}, \citenamefont {Saive},
  \citenamefont {Fr{\'e}d{\'e}rick}, \citenamefont {Bichler}, \citenamefont
  {Amann}, \citenamefont {Holleitner}, \citenamefont {Kaniber},\ and\
  \citenamefont {JJ}}]{Laucht2012Single}%
  \BibitemOpen
  \bibfield  {author} {\bibinfo {author} {\bibfnamefont {A.}~\bibnamefont
  {Laucht}}, \bibinfo {author} {\bibfnamefont {S.}~\bibnamefont {P{\"u}tz}},
  \bibinfo {author} {\bibfnamefont {T.}~\bibnamefont {G{\"u}nthner}}, \bibinfo
  {author} {\bibfnamefont {N.}~\bibnamefont {Hauke}}, \bibinfo {author}
  {\bibfnamefont {R.}~\bibnamefont {Saive}}, \bibinfo {author} {\bibfnamefont
  {S.}~\bibnamefont {Fr{\'e}d{\'e}rick}}, \bibinfo {author} {\bibfnamefont
  {M.}~\bibnamefont {Bichler}}, \bibinfo {author} {\bibfnamefont {M.-C.}\
  \bibnamefont {Amann}}, \bibinfo {author} {\bibfnamefont {A.}~\bibnamefont
  {Holleitner}}, \bibinfo {author} {\bibfnamefont {M.}~\bibnamefont {Kaniber}},
  \ and\ \bibinfo {author} {\bibfnamefont {F.}~\bibnamefont {JJ}},\ }\href
  {\doibase 10.1103/PhysRevX.2.011014} {\bibfield  {journal} {\bibinfo
  {journal} {Physical Review X}\ }\textbf {\bibinfo {volume} {2}},\ \bibinfo
  {pages} {011014} (\bibinfo {year} {2012}{\natexlab{b}})}\BibitemShut
  {NoStop}%
\bibitem [{\citenamefont {Rao}\ and\ \citenamefont
  {Hughes}(2007{\natexlab{b}})}]{Rao2007Beta}%
  \BibitemOpen
  \bibfield  {author} {\bibinfo {author} {\bibfnamefont {V.~M.}\ \bibnamefont
  {Rao}}\ and\ \bibinfo {author} {\bibfnamefont {S.}~\bibnamefont {Hughes}},\
  }\href {\doibase 10.1103/PhysRevB.75.205437} {\bibfield  {journal} {\bibinfo
  {journal} {Physical Review B}\ }\textbf {\bibinfo {volume} {75}},\ \bibinfo
  {pages} {205437} (\bibinfo {year} {2007}{\natexlab{b}})}\BibitemShut
  {NoStop}%
\end{thebibliography}%

\end{document}